\documentclass[prb,showpacs,preprintnumbers,amssymb,twocolumn]{revtex4}
\usepackage{graphicx}
\usepackage{dcolumn}
\usepackage{bm}
\def\be{\begin{equation}}
\def\ee{\end{equation}}

\begin{document}

\title{Distribution of conductance for Anderson insulators: A theory with a single parameter}
\author{Andrew Douglas and K. A. Muttalib}
\affiliation{Department of Physics, University of Florida, P.O. Box 118440,
Gainesville, FL 32611}
\begin{abstract}
We obtain an analytic expression for the full distribution of conductance for 
a strongly disordered three dimensional conductor within a perturbative approach based on transfer matrix formulation. Our results confirm the numerical evidence that the log-normal limit of the distribution is not reached even in the deeply insulating regime. We show that
the variance of the logarithm 
of the conductance scales as a fractional power of the mean, while the skewness changes sign as one approaches the Anderson metal-insulator transition 
from the deeply insulating limit, all described as a function of a single parameter. The approach suggests a possible single parameter 
description of the Anderson transition that takes into 
account the full non-trivial distribution of conductance. 

\end{abstract}


\maketitle

Quantum fluctuations have been intensely studied in recent years, but in many 
cases a fundamental understanding of their effects on physical observables remain 
poorly understood. In particular the effects of large mesoscopic fluctuations on 
quantum phase transitions, both in interacting electron systems \cite{loh} and in 
disordered non-interacting models  \cite{lee}, have not been studied systematically.
For a non-interacting  system, while the  distribution $P(g)$ of the 
dimensionless conductance $g$ in the metallic weak disorder limit is well understood
\cite{stone}, even a qualitative understanding of $P(g)$ is lacking in the 
strongly disordered insulating regime in three dimensions (3D) where numerical data 
show large deviations \cite{markos} from an expected log-normal distribution.
The more fundamental question  
of how this non-trivial distribution changes as one 
decreases the disorder from the deeply insulating regime toward the Anderson 
metal-insulator transition point has remained largely unexplored. This is primarily due to 
the lack of appropriate theoretical tools to consider such distribution functions 
analytically. Conventional field theory framework 
\cite{shapiro} which relies on a small $\varepsilon$ expansion in $2+\varepsilon$ 
dimensions gives results that do not agree even qualitatively, in the $\varepsilon \rightarrow 1$ limit, 
with numerical results in 3D \cite{soukoulis}. 

In this work, we obtain the full conductance distribution in 3D in the 
strong disorder regime analytically within a perturbative approach, and show that 
the results are consistent with available numerical 
data. Our model involves a single disorder parameter $\Gamma=\xi/4L_z$ where $\xi$ is 
the localization length and $L_z$ is the length of the conductor. A second parameter
$\gamma =\xi/8L$, where $L$ is the cross-sectional dimension, becomes independent only if arbitrary geometrical shapes are considered.
We show first of all that the interaction between different channels of a conductor 
in 3D remains important even in the deeply insulating regime; the result is that 
the 3D distribution is never log-normal. Instead, we find that the variance of the logarithm of conductance scales approximately as $2/3$rd power of the mean
\cite{somoza}.
In addition, we also explicitly evaluate the third cumulant 
$\kappa_3 \equiv \langle (\ln g-\langle \ln g \rangle )^3\rangle$   
which describes the asymmetry of the distribution. We find that in the deeply insulating 
regime $\kappa_3 $ is positive (longer tails toward larger conductances). It then decreases 
with decreasing disorder according to $\kappa_3 \sim \langle -\ln g \rangle$, going through 
zero well before the Anderson metal-insulator transition point \cite{critical}. As one 
decreases the disorder further toward the critical point, $\kappa_3$ becomes negative, 
describing increasing asymmetry in the opposite direction \cite{vR}. We emphasize that all of this 
is described as a function of a single parameter that fixes the mean value. 
Our method therefore allows us for the first time to 
explore in quantitative detail how the conductance distribution changes as one approaches 
the Anderson transition point starting from the deeply insulating regime, and how it can 
still be described analytically within a one parameter theory.

Our formulation is based on the transfer matrix framework developed originally for transport 
in quasi one-dimension (Q1D) where the transverse length of a conducting wire is 
less than the localization length. In Q1D, the Dorokhov-Mello-Pereyra-Kumar (DMPK) 
equation \cite{dmpk} has been enormously successful in describing the details of the 
distribution of the transmission eigenvalues \cite{been-rmp}. 
Exploiting the Landauer formula 
\cite{landauer} to connect the distribution of $g$ with the distribution of the 
transmission eigenvalues, the DMPK equation has been used  to 
obtain a variety of novel features in the distribution of conductances in Q1D \cite{mwg}.
A generalization of the Q1D DMPK equation, claimed 
to be valid in 3D, has been proposed in [\onlinecite{mk}]. By solving this so called 
Generalized DMPK (GDMPK) equation numerically and comparing the results with those 
from direct numerical solution of the tight binding Anderson model, it has recently 
been shown \cite{brn} that the GDMPK not only incorporates the effects of 
dimensionality correctly, but that it also describes the full distribution of the 
transmission levels quantitatively 
in the insulating as well as near the critical regime in 3D. 
However, 
analytic solutions of the equation in the strongly disordered regime have been 
obtained only within a very approximate saddle point scheme, where the interaction 
between the transmission eigenvalues are entirely ignored \cite{mmwk}. 
These approximate solutions do show deviations from the Q1D behavior in the right 
direction, but they fail to describe correctly how e.g. the variance or the skewness of the distribution changes with disorder. 

Here we solve 
the GDMPK equation analytically, including for the first time the interaction between the eigenvalues,  
within a novel perturbative approach. We find that the density of the transmission 
eigenvalues in the insulating regime is a constant, with 
an exponential gap at the origin that increases with increasing disorder. This is fundamentally different from the approximate solutions where interaction between eigenvalues are neglected. The resulting analytic 
expression for the full $P(\ln g)$ as a function of disorder now agrees \textit{quantitatively} with 
available numerical data. The results suggest a possible single parameter description of the Anderson 
transition starting from the insulating side and taking into account the full distribution 
of conductances even when the average (or the most probable value) of the conductance is no longer a 
meaningful representation of a highly non-trivial asymmetric distribution.   

For an $N$-channel disordered conductor of fixed cross section $L^2$, the GDMPK equation \cite{mk} describes the evolution with
length $L_z$ of the joint probability distribution $p_{L_z}(\textbf{x})$ of the $N$ transmission 
eigenvalues $x_i$:
\begin{eqnarray}\label{mg}
\frac{\partial p(\textbf{x},t)}{\partial t} = \frac{1}{4} 
\sum_{i=1}^N \frac{\partial}{\partial x_i}K_{ii}\left[ \frac{\partial}{\partial x_i}+
\frac{\partial \Omega}{\partial x_i}\right]p(\textbf{x},t) \cr
\Omega  \equiv  -\sum_{i<j} \gamma_{ij} \ln | f(x_i,x_j) | - \lambda
\sum_i \ln |\sinh 2x_i |
\end{eqnarray}
with the initial condition $p(\textbf{x},t=0)=\delta(\textbf{x})$, where 
$t\equiv L_z/l$ with $l$ being the mean free path, $f(x_i,x_j)\equiv \sinh^2 x_j-\sinh^2 x_i$ and we have kept $\lambda=1$ as a 
free parameter for later convenience.
Here $\gamma_{ij}\equiv 2K_{ij}/K_{ii}$, where $K_{ij}$ is a phenomenological matrix
defined in terms of certain eigenvector correlations that can be explicitly evaluated
numerically. The Q1D DMPK equation 
is recovered when $\gamma_{ij}=1$ (we only consider orthogonal symmetry). In Ref~[\onlinecite{brn}] it was shown that only two parameters, $K_{11}$ and $K_{12}$, are enough to model the entire matrix $K_{ij}$ in the insulating as well as the critical regimes, as proposed in Ref~[\onlinecite{mmwk}]. 
Disorder is then characterized by the parameter $\Gamma \equiv l/K_{11}L_z$.
In the insulating side, one can interpret $\xi\equiv 4l/K_{11}$ as the localization 
length, so $\Gamma = \xi/4L_z$. The other parameter
$\gamma_{12}\equiv \gamma= \xi/8L$, so that $\Gamma/\gamma=2L/L_z$ depends on geometry. 
The entire distribution, for a cubic system, is therefore
characterized by a single disorder parameter $\Gamma$ (with $\gamma=\Gamma/2$).

We note that it is through the $L$ dependence of the eigenvector correlations $K_{11}$ and $K_{12}$ that the GDMPK `knows' about the dimensionality of the system. In 3D, $K_{11}\sim 1/L^0$ ($1/L^2$) in the insulating (metallic) regime; the quantity $\tilde{K}_{11}\equiv \lim_{L\rightarrow \infty}K_{11}(L)$ is zero in the metallic regime as well as at the critical point, but is finite for insulators \cite{mmwk}. In 2D on the other hand, $\tilde{K}_{11}$ is always finite. Thus the particular $L$ dependence of $K_{11}$, and therefore of our parameter $\gamma$, not only reflects the proper dimensionality, but also contains information about the critical point.

Once an analytic  
solution of the GDMPK equation for $p(\textbf{x})$ is available, the full distribution of 
conductances $P(g)$ can be obtained via the Landauer formula\cite{landauer,mu-wo} 
\be \label{landauer}
P(g)\propto \int\prod_a^N dx_a p(\textbf{x})\delta
\left(g-\sum_i \rm{sech}^{2}x_i\right).
\ee 
We can therefore expect that Eq.~(\ref{mg}) can be used as the
starting point for studying the distribution of conductances at large disorder. Of course the DMPK itself breaks down for $\xi < l$; however this happens far from the transition region and we do not consider such extreme disorder.

We first briefly outline our method used to solve Eq.~(\ref{mg}) in the insulating regime. 
Following Ref.~[\onlinecite{beenakker}] we use a factorization in terms of $\zeta\equiv e^{-\Omega/2}$:
\begin{eqnarray} \label{singular}
p(\textbf{x},t) = \lim_{\textbf{y}\rightarrow 0} \zeta(\textbf{x})G_N(\textbf{x};t|\textbf{y})\zeta^{-1}(\textbf{y})  \cr
= \lim_{\textbf{y}\rightarrow 0}\prod_{i<j}
\frac{|f(x_i,x_j)|^{\frac{\gamma}{2}}}
{|f(y_i,y_j)|^{\frac{\gamma}{2}}}
\prod_i \frac{|\sinh 2x_i|^{\frac{\lambda}{2}}}{ |\sinh 2y_i|^{\frac{\lambda}{2}}}
 G_N(\textbf{x},t|\textbf{y}).
\end{eqnarray}
The GDMPK transforms into an equation for the $N$-particle Greens function $G_N$:
$
-\partial G_N/\partial t=HG_N
$, with
 
\begin{eqnarray}  \label{ham}
&H &= -\frac{1}{2\tilde{\gamma}}\sum_i \frac{\partial^2}{\partial x^2_i} 
 +  \frac{\lambda(\lambda-2)}{2\tilde{\gamma}} \frac{1}{\sinh^2 2x_i} \cr 
 &+& \frac{\gamma(\gamma-2)}{4\tilde{\gamma}}\sum_{i < j}  \left[\frac{1}{\sinh^2(x_i-x_j)}+\frac{1}{\sinh^2(x_i+x_j)}\right]
\end{eqnarray}
where $\tilde{\gamma}\equiv \xi/4l$. 
The initial condition is
$G(\textbf{x};t=0|\textbf{y}) = \frac{1}{N!}\sum_{\pi(\textbf{y})}\delta(\textbf{x}-\textbf{y})$
where $\pi(y)$ refer to a symmetrized permutation. This maps the problem onto a set of $N$ interacting bosons evolving in imaginary time in 1D with delta function initial conditions \cite{note}. 

In the strongly disordered limit the interaction strength $\gamma \ll 1$, and we exploit this small parameter to develop a perturbation theory to evaluate the non-equilibrium Green’s function using standard Keldysh techniques \cite{mahan}.  However, the small $\textbf{y}$ singular behavior in the denominator of Eq.~(\ref{singular}) demands that $G_N(\textbf{x},t|\textbf{y})$ must go as $\sim \zeta(\textbf{y})$ in the small $\textbf{y}$ limit in order to recover a well-defined $p$.  To extract this non-analytic behavior of $G_N$ from the diagrammatic expansion, we must sum the expansion in a particular way. 
First we treat $\tilde{\lambda}\equiv (\lambda -2)$ in Eq.~(\ref{ham}) as a small parameter in which we will later set $\lambda = 1$.  This is justified because the short range single particle potential primarily serves to provide a boundary condition at the origin and its actual strength turns out to be unimportant. Next we expand the Green’s function in a Taylor series in both $\gamma$ and $\tilde{\lambda}$. 
Upon reorganizing the Taylor series expansion of $G_N$ into an exponential series, 
we can factor out from the series expansion
the unperturbed $N$-particle Green’s function, the exact single particle Green’s function $G_1(i)$ (defined through Eq.~(\ref{ham}) with $\gamma = 0$) and exact two particle Green’s function $G_2(i,j)$ (defined through Eq.~(\ref{ham}) with $\tilde{\lambda}=0$).  Keeping just these terms, and neglecting those terms remaining in the series expansion results in our `first order' approximation to the full $G_N$,
\be
G_N(t)\approx G^0_N(t) \prod_i\frac{G_1(i)}{G_1^0(t)}\prod_{i<j}\frac{G_2(i,j)}{G_2^0(i,j)},\label{GN}
\ee
where the superscript $0$ refers to the unperturbed solution. These turn out to be the dominant terms for all values of $\textbf{x}$, $\textbf{y}$, and in particular they reproduce the non-analytic behavior aforementioned. Note that the approximation is `first order' in the sense that the terms neglected in the series are all of order $\gamma^2$ or higher; all terms of order $\gamma$ are kept, and some of the terms are summed up to infinite order in $\gamma$.
The approximation turns out to be quite good even close to the Anderson transition where $\gamma$ remains smaller than unity, but clearly it can be improved systematically by repeating the above procedure for higher orders in $\gamma$.

In order to obtain explicit results, we evaluate $G_1(i)$ and $G_2(i,j)$ in the limit $\gamma \ll 1$:
\begin{eqnarray}\label{G}
\frac{G_1(x;t|y)}{G^0_1(x;t|y)} &\approx & x \tau(2y) \cr
\frac{G_2(x_1,x_2; t|y_1,y_2)}{G^0_2(x_1,x_2,t;|y_1,y_2)} 
&\approx & T_{\gamma}(x_1+x_2)T_{\gamma}(|x_1-x_2|) \cr
&\times& \tau(y_1+y_2)\tau(|y_1-y_2|)
\end{eqnarray}
where
$\tau(z)\equiv \tanh^{\frac{\gamma}{2}}z$ and 
\be
T_{\gamma}(x) \equiv \left[1-\frac{\gamma}{\sqrt{2\Gamma}} \rm{erfc'}\left(\frac{\gamma}{2\sqrt{2\Gamma}}+\frac{x\sqrt{2\Gamma}}{2}\right) \right] \tau(x).
\ee
Here $\rm{erfc'}(x)\equiv (\sqrt{\pi}/2)e^{x^2}\rm{erfc}(x)$ where $\rm{erfc}(x)$ is the complementary error function.
As a first check, we have verified that our solution for the distribution of the transmission levels obtained from Eq.~(\ref{singular}) agrees  with exact solutions known in the Q1D limit \cite{beenakker} for the special values of $\gamma=1,2$ and $4$, for both small and large $x$ where simple analytic expressions are available. In order to solve for the distribution of conductances using Eq.~(\ref{landauer}), we consider the free-energy of $N$ interacting particles on a line. First of all, we obtain the density of the eigenvalues $\sigma(x)$ as a function of the disorder parameter $\Gamma$:   
\be \label{density}
\int_0^{\sigma}\frac{d\sigma'}{\sigma'\sqrt{2\gamma\sigma'-4\Gamma \ln [\sigma' e/4}]}=x.
\ee
The small and large $x$ behavior are given by 
$\sigma(x)  \approx  \frac{2\Gamma'}{e\gamma}e^{-\Gamma'(x-x_{sp})^2}$ for $x \ll x_{sp}$, and 
$\sigma(x) \approx  \frac{2\Gamma'}{\gamma}$ for $x \gg x_{sp}$. Here $x_{sp}\approx [1/2\Gamma +1]$
and $\Gamma'\approx [\Gamma - 2\Gamma^2]$. Note that $\Gamma \ll 1$ in the insulating regime. 
Figure 1 shows the density according to Eq.~(\ref{density}) for two different values of $\Gamma$, with exponential gap at the origin that increases with increasing disorder. 
\begin{figure}
\includegraphics[angle=0,width=0.40\textwidth]{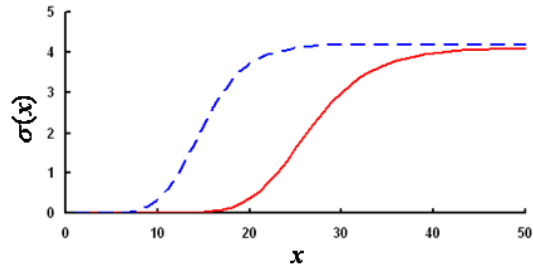}
\caption{(Color online) Density of the transmission eigenvalues for two values of disorder, $\Gamma = 0.014$ (solid red line) and $\Gamma=0.0285$ (dashed blue line). The exponential gap at the origin increases with increasing disorder (decreasing $\Gamma$).}
\end{figure}
In contrast, the density in the metallic regime is also constant, but starting at the origin  \cite{been-rmp}. Thus our result suggests that the opening of a gap in the eigenvalue spectrum could be considered as a signature of the metal-insulator Anderson transition.

The conductance in the strongly disordered regime $\gamma \ll 1$ is dominated by the smallest eigenvalue $x_1$, although in contrast to Q1D it remains highly interacting. In this regime the distribution can be expressed in a simple form: 
\begin{eqnarray}\label{P}
 P(\ln g)& = & \exp\left[-f\left(\frac{1}{2}\ln \frac{4}{g}\right)\right]; \cr
f(x) & \equiv & v(x)- \frac{\sqrt{\pi}}{8e\gamma\sqrt{\Gamma'}} \rm{erfc}\left[(x-x_{sp})\sqrt{\Gamma'}\right]
\end{eqnarray}
where 
$v(x)\equiv [\Gamma x^2-\ln x-\frac{1}{2}\ln \sinh 2x]$. 
Figure 2 shows plots for $P(\ln g)$ obtained from Eq.~(\ref{P}) for three values of disorder, compared with numerical data 
\begin{figure}
\includegraphics[angle=0,width=0.45\textwidth]{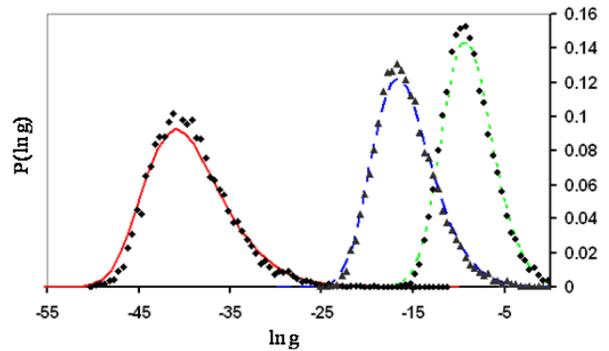}
\caption{(Color online) $P(\ln g)$ in the insulating regime for three values of disorder, $\Gamma = 0.014$ (solid red line), $\Gamma=0.0285$ (dashed blue line) and $\Gamma=0.051$ (dotted green line) corresponding to $\langle \ln g \rangle = -39.4$, $\langle \ln g \rangle = -15.8$ and $\langle \ln g \rangle = -8.9$, respectively. The numerical data points are from Ref~[\onlinecite{pm}], for the same values of $\langle \ln g \rangle $.}
\end{figure}
obtained from solving the tight binding Anderson model. In order to compare the results in more detail, we plot in Figure 3 the variance $\sigma_2$ of $\ln g$, which is consistent with a power law $2/3$rd \cite{somoza}.
Figure 4 shows the third cumulant $\kappa_3$ which behaves as $\kappa_3 \sim \langle -\ln g \rangle$, and in the inset we plot the corresponding skewness $\chi\equiv \kappa_3/\sigma_2^{3/2}$ which seems to saturate at a value $\sim 0.8$.
Note that the skewness changes sign around $ \langle -\ln g \rangle \approx 5$, which is still far from the Anderson transition which occurs \cite{critical} around  $ \langle -\ln g_c \rangle \approx 1.3$. Thus the distribution changes its shape from a large positive asymmetry to a negative one that increases as the disorder is decreased toward the critical point. 

\begin{figure}[ht]
\includegraphics[angle=0,width=0.45\textwidth]{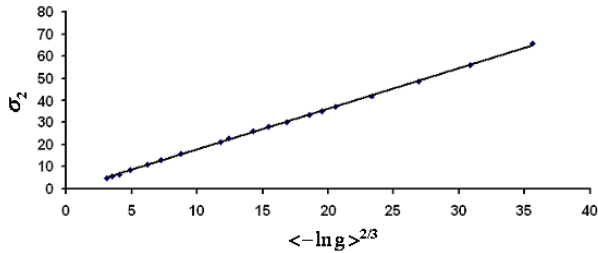}
\caption{variance of $\ln g$ plotted as a function of $\langle -\ln g \rangle^{2/3}$. 
The points are calculated from Eq.~(\ref{P}), and the line is a fit. }
\end{figure} 
\begin{figure}
\includegraphics[angle=0,width=0.45\textwidth]{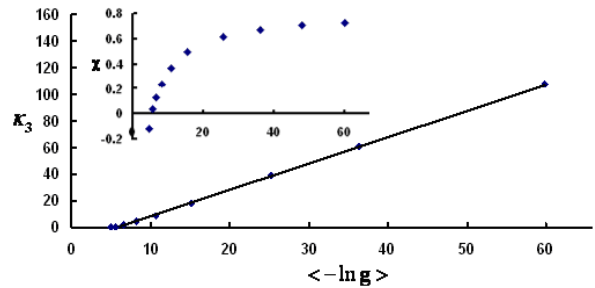}
\caption{The third cumulant $\kappa_3$ as a function of $\langle -\ln g \rangle$. The inset shows the corresponding skewness $\chi$. Note that $\chi$ (or $\kappa_3$) changes sign at $\langle -\ln g \rangle \approx 5$. The Anderson transition occurs at $\langle -\ln g \rangle \approx 1.3$.} 
\end{figure} 

How close to the critical point can we approach with our present formulation? It turns out that once the skewness becomes negative, the contributions to the conductance from eigenvalues other than the smallest one become important. 
Nevertheless, the value of the variance at the critical point $\langle -\ln g \rangle$ is $1.64$, close to $1.09$ obtained from numerical data \cite{critical}. This suggests that it should be possible to improve the calculations systematically to study the conductance distribution at the critical point
by increasing the number of eigenvalues included in the calculation of the conductance distribution. 

In summary, we have developed a perturbative approach that allowed us to obtain, from a transfer matrix formulation, the full distribution of conductances in the insulating regime of a 3D disordered conductor. The solution takes into account the interaction between the transmission eigenvalues that had been ignored in the past. The formulation involves a phenomenological matrix characterizing eigenvector correlations; analyzing the properties of this matrix numerically allowed us to consider a simplified model with two independent matrix elements only.
The distribution is then obtained as a function of a single disorder parameter that fixes the mean value $\langle \ln g \rangle$, 
even though 
$P(\ln g)$ changes its shape from a positive to a negative skewness as the disorder is decreased from the deep insulating regime toward the critical Anderson transition point. The results agree with recent numerical simulations of the full distribution.
With current fabrication technology, it should be possible to verify the predictions experimentally. While the method is developed for the insulating phase only, 
it also leads to a possible characterization of the Anderson transition in terms of the opening of a gap in the spectrum of the transmission eigenvalues. By systematically improving the approximations, the method could therefore be used to study the qualitative features of the critical distribution near the Anderson transition point.  

We gratefully acknowledge helpful discussions with F. Evers and P. Marko\v{s}.


\begin{thebibliography}{99}
\bibitem{note} One could equivalently choose to work with $N$ fermions by choosing $\zeta$ to be antisymmetric \cite{been-rmp}.  The only restriction is that $p$ must be symmetric.
\end{thebibliography}
\end{document}